\documentclass[twocolumn,showpacs,preprintnumbers,amsmath,amssymb]{revtex4}
\usepackage{graphicx}
\usepackage{dcolumn}
\usepackage{bm}

\newcommand{\bi}{\begin{itemize}}
\newcommand{\ei}{\end{itemize}}
\newcommand{\be}{\begin{equation}}
\newcommand{\ee}{\end{equation}}
\newcommand{\ba}{\begin{eqnarray}}
\newcommand{\ea}{\end{eqnarray}}
\newcommand{\bse}{\begin{subequations}}
\newcommand{\ese}{\end{subequations}}

\begin{document}

\title{Covariant description of spin $1/2$ particles in self--action} 

\author{Yair Goldin$^\dagger$}

\email{yair@nuclecu.unam.mx}

\altaffiliation[Postal address: ]{Lomas de Tarango 155, 
M\'exico D.F.,  01620, M\'exico} 

\affiliation{$^\dagger$Facultad de Ciencias\\ Universidad Nacional Aut\'onoma de
M\'exico\\ Circuito Exterior C.U., M\'exico D.F.,  04510, M\'exico}

\begin{abstract}
The bispinor wave function finds its fundamental application in the study of
electrons, neutrinos and protons as particles bound by their own potentials.
Classical electromagnetism and the Dirac electron theory appear to be natural
extensions of the physics that rules the electron itself. However, the charge
density and the negative energy eigenvalues are two concepts which cannot
survive further scrutiny. The electron and the proton resemble minute planetary
systems where the bispinor terms defining their physical properties revolve with
definite angular momentum around the singular point of the particles' potentials.
The mass of the electron--neutrino in relation to the electron's and the mass of
the latter in relation to the proton's are determined under particular assumptions.
The most important feature of the new approach is the faithful representation of
free particles at rest with respect to inertial systems of reference. This certainly
enhances the foundation of physics; precisely in the realm of Quantum Mechanics,
where the customary description of free electrons is a confusing consequence of the
limited scope of the Schr\"{o}dinger and Dirac equations.
\end{abstract}


\maketitle
 
\section{General considerations}
The free particle solutions of the
Schr\"{o}dinger equation, plane waves  extending throughout space,
bear no direct relation with the concept of particle that most of us may have. It is
only natural that this circumstance gave rise to philosophical currents and
crosscurrents from which no real consensus was ever obtained: on examining the
probabilistic interpretation of the wave function,  Niels Bohr arrived at the
conclusion that science not necessarily should be concerned on whether or not the
dynamic variables of the electron coexist with definite values. In essence, he
argued that physical reality cannot be dissociated from the process of measurement.
Consequently free electrons, which by definition do not interact even with
measuring devises, remain metaphysical until their state of isolation is disrupted
by experiment. Moreover, the Heisenberg uncertainty relations have revealed  an
intrinsic limitation in the accuracy that can be attained when the
position and the momentum of the electron are measured simultaneously. Hence the
entire formalism of Quantum Mechanics is to be considered as the tool for deriving
predictions of statistical nature~\cite{[1]}

However, contrary to
the familiar answer of the Copenhagen school, perhaps in accordance with
Einstein's faith in an objective reality, physics is sustained on the
principle of inertia, which  does not allow room for indeterminacies of  the kind.
For instance, the parameter of relative velocity appearing in the Lorentz
transformations implies the presence of  a particle at rest with respect to  one of
the inertial systems, and in uniform motion with respect to the other. Why? 
Because the relative velocity between systems of coordinates is devoid
of physical meaning until a real object (particle) appears in space. We may ask, 
what sense does it make to verify the Lorentz covariance of the Dirac equation
when the equation itself cannot describe a particle in isolation in accordance with
the principle that makes the Lorentz transformation meaningful: the Galileo--Newton
principle of inertia in the first place.

As a matter of fact, Erwin Schr\"{o}dinger solved the hydrogen atom thinking of a
proton fixed at the origin of an inertial system implied by the
potential $r^{-1}$ in the equation. If the electron and the proton
revolve around the center of mass, it is irrelevant for the point in question: 
it only means that in quantum mechanics one assumes that the atom remains fixed
with respect to an inertial system. It follows, therefore, that in the attempt
to preserve its conceptual consistency, quantum mechanics had to invoke the
process of measurement or to rely on some other interpretation.  Physics evolved
leaving behind important conceptual problems unsolved. And so, some twenty years
after the advent of the Schr\"{o}dinger equation this problem emerged with a
different face.  Physicists became fully aware that, in order to
complete the theory of the free electron, self--action had to be reckoned with.
Then QED offered an approach based on the idea that the point--electron emits and
absorbs virtual and real photons along its path. This time, however, the
divergent self--energy put a definite end to the hope of a consistent theory, not
to mention that the QED approach itself was inconsistent with the principle of
inertia from the very outset, since the free electron path cannot be made to vanish
with the choice of system of reference~\cite{[2]}.

       Free electrons exist independently of experiment; we think that it is
of the concern of physics to explain their existence in a way consistent with
general principles. Fortunately, the framework to analyze the problem is defined
already:  the description of a self-interacting electron at rest with respect to an
inertial system of reference should be carried out with the solutions of the one
and only fundamental equation that can be proposed for the study of spin $1/2$
particles in self-action:
\begin{equation}\gamma _\mu \left[ {\partial _\mu +S_\mu (e,\nu, p )} \right]\,\psi
\ = \ 0,\label{eq1}\end{equation}
where $S_\mu$ represents the self--action 4--vector of the particle under study. 

The conceptual structures of equation (1) and of the Dirac equation
\cite{[3]}, 
\begin{equation}  \gamma_\mu\,\left[\partial_\mu
+\alpha\,A_\mu\right]\,\psi \ = \
-\,i\,\left(m\,c\,\hbar^{-1}\right)\,\gamma_{_5}\,\psi,\label{eq2}\end{equation}
may reinforce each other in their own range of applicability and may fuse
at  some point too. The free particle solutions of the customary wave equations are
important, but only as part in the solution of problems with discontinuous external
interaction, either in space or in time. The tunnel effect --through the potential
barriers that limit the  force free region of the potential well-- is a typical case
where plane waves satisfying boundary conditions become necessary and meaningful
also. Although, we suspect that plane waves with no boundary conditions are
meaningless solutions which have misled physicists for over 75 years now.

\section{The electron self--potentials}

Mostly every physicist knows that the energy of a point electron diverges;
although not every physicist is well aware that Gauss' law cannot hold when
the particle is deprived of physical dimensions. Indeed, the volume integral
of the last two expressions below reveals a contradiction:
\begin{equation} \nabla \cdot \vec{\textrm{E}} \ = \ r^{-2}\partial _r(r^2\,
\hat{\textrm{r}}\cdot \vec{\textrm{E}}) \ = \ r^{-2}\partial _r(1) \ = \ 4\,\pi
\,\delta (r).\label{eq3}
\end{equation}
The reason some textbooks invoke the divergence theorem to \textit{prove}
(\ref{eq3}) correct is  easy to understand: empty space cannot be the origin of the
Coulomb potential. Interesting enough, this is not the only case where physical
problems have been  circumvented at the expense of mathematical congruence. Paul
Dirac himself had the opinion that renormalization has defied all the attempts of
the mathematician to make it sound \cite{[4]}; yet, such procedures have advocates
and detractors who may never convince one another.

        It is appropriate to recall that the functional dependence of the 
potential  on the charge density brings about problems of its own. For example,
who can tell if the electrostatic energy of an electron is localized where the
potential and the charge density coexist or is it dispersed throughout space?
\begin{equation} {\cal E} \ = \
\int\limits_{\textrm{\tiny{electron}}}{\rho\,\phi}  \ = \
\int\limits_{\textrm{\tiny{all space}}}{\nabla\phi\cdot \nabla\phi}.
\label{eq4} \end{equation}
More important, the association of the square of the electrostatic field with  
the energy density leads to the following conclusion: the ratio of
electromagnetic momentum over electromagnetic energy  of  an electron in uniform
motion is incompatible with relativity \cite{[5]}. 

Experience shows that each and every electron in a given distribution 
contributes with their individual Coulomb potential  to the total potential.
However, if the electron happens to be a particle with physical dimension,
there is no reason to think that a small  portion  of its volume would
contribute with a small portion to  its Coulomb potential. Again, since there
is no consistent model of the electron,  the physics that make its existence
possible is  not known.

    The vanishing of the Laplacian of the inverse distance occurs in three 
dimensions only:
\begin{equation}\nabla^2_{\textrm{N}}\left(\sum_{k\,=\,1}
^{\textrm{N}}{x_k^2}\right)^{-1/2}
\ =
\ 0
\qquad \Rightarrow\qquad \textrm{N} \ = \ 3.\label{eq5}
\end{equation}
In fact, this is a mathematical characteristic  of the Coulomb potential
inherent to the electron. This is a postulate and it is incompatible with the
concept of charge density. 

The task now is to adjust the electromagnetic theory to
a  new conceptual  structure. We shall do that in parts and as the circumstance
allows.
  It  is convenient to say in advance that suitable combinations of the products 
of the four components of the wave function and their complex conjugates enable 
the construction of covariant densities. These densities play an essential role in
the definitions of the electromagnetic inertia  and of the other properties of the
electron; all of which properties  will vanish beyond a small radius from the
singular point of the potential (from now on ``singularity'' for short). In this way
the Coulomb potential springs  up from the inside of the electron; though, 
no functional relation with any of the covariant densities exists.  

Electromagnetic signals from the singularity necessarily relate to the 4-vector of
null length, $r_\alpha=(\vec{\textrm{r}},r)$. The contraction
of $r_\alpha$ with the 4-velocity of the singular point, 
\begin{equation} u_\alpha \ = \
(\vec{\textrm{u}},1)\,(1-\textrm{u}^2)^{-1/2}\ \to\,
(\vec{\textrm{0}},1),\label{eq6}\end{equation}
yields the fundamental invariant involved in the Lienard-Wiechert potentials 
\ba \textrm{I}_0 && = \ -(r_\alpha\,u_\alpha)^{-1}\,\to\,r^{-1},\label{eq7}\\
A_\alpha && = \ \textrm{I}_0\,u_\alpha \ = \ 
(\vec{\textrm{u}},1)\,(r-\vec{\textrm{r}}\cdot \vec{\textrm{u}})^{-1}\
\to\ (\vec{\textrm{0}},r^{-1}),\nonumber\\
\label{eq8}\ea
(the arrow reads --for a particle at rest reduces to--). 

Direct consequence of 
their genesis, these potentials satisfy the fundamental  differential
equations
\ba  \partial_\alpha\,\partial_\alpha\,A_{\beta} \ = \
0,\label{eq9}\\ \partial_\alpha\,A_\alpha \ = \ 0,\label{eq10}\ea
Consequently, the field tensor $F_{\alpha\beta}=\partial_\alpha\,A_\beta -
\partial_\beta\,A_\alpha $ satisfies Maxwell's equations with no sources: 
\ba
\partial_\alpha\,F_{\beta\gamma}+\partial_\gamma\,F_{\alpha\beta}+\partial_\beta\,
F_{\gamma\alpha} \ = \ 0,\label{eq11}\\
\partial_\alpha\,F_{\alpha\beta} \ = \ 0,\label{eq12}\ea
In order to write in a concise manner the potentials of a continuous
distribution of electrons in arbitrary motion, one may assume that the customary
equations,
$\partial_\alpha\,\partial_\alpha\,A_\beta=-J_\beta$ and
$\partial_\alpha\,A_\alpha=0$,  hold. However, in this case $J_\beta$  would represent the
piecewise differentiable functions best approximating the 4-current of singularities. The
vanishing of the 4--divergence\, $\partial_\alpha\,J_\alpha$ \, states the
conservation of the net number of singular points.

    The repulsive Coulomb potential alone is not sufficient to get square
integrable solutions.  Tentatively, the second potential that is necessary for
this purpose could be thought of as the  quantum version of the vector potential of
a point magnet:  
$ (\lambda_e/2)\,(\vec\sigma \times
\vec{\textrm{r}}\,\,r^{-3})$,  where 
$\vec\sigma$    represents the $4\times 4$ Pauli  spin matrices, and \, $
\lambda_e$ \, is a length--parameter measuring the strength of the coupling,
not to be fixed before hand. The reader may wish to verify that the
substitution of the matrix potential in equation (\ref{eq1}) is entirely
equivalent to the substitution of the imaginary vector  $-\,i\,\lambda_e
r^{-2}\,\hat r$. The covariant form of this 3--vector enters as part of the
self--action 4--vector: 
\ba S_\mu(e)=-\alpha\,
A_\mu+i\,\lambda_e\sigma_\mu
=-\alpha \, \textrm{I}_0\,u_\mu+i\,\lambda_e\partial_\mu\, \textrm{I}_0\nonumber\\
\to\qquad -\alpha\,(\vec{\textrm{0}},r^{-1})-i\,\lambda_e(r^{-2}\hat
r,0)\label{eq13}\ea
 
 The electromagnetic coupling constant in the equation above means that the
electron was assumed to repel itself with the same strength that two electrons
repel each other. Now, the imaginary radial potential is also indispensable for the
description of the electron--neutrino. As neutrinos may not have magnetic
moment, perhaps this gauge invariant
($\sigma_{\alpha\beta}=\partial_\alpha\sigma_\beta
-\partial_\beta\sigma_\alpha=0$) new entity in physics is not related to magnetic
moment but to the spin. For simplicity,  the coupling constant
$\lambda_e$ was considered to be the same for both particles. Eventually we  will
relate 
$\alpha$   to 
$\lambda_e$, which in turn could relate $\alpha$ to $m_\nu/m_e $ if $\lambda_e$ is
written in terms of $m_e,\,m_\nu,\,\hbar$ and $c$. There are  two options:
\begin{equation}\lambda_e \ = \ \frac{\beta\,\hbar}{m_e
\,c},\label{eq14}\end{equation} 
where $\beta=m_\nu/m_e$. Interchanging the mass parameters would be the second
option, but it does not work.

\section{The electron solutions}

An electron in self--action is a state with definite energy which is taken into
account with the factor $\exp(-i\,E\,t/\hbar) $ in the solutions. Since the
potentials are radial, the spatial part of  the two independent solutions is to
be written as shown~\cite{[3]}.\\

\noindent
First solution:
\ba \psi_1 &&= 
\left[\frac{j+1-m}{2(j+1)}\right]^{1/2}F\,Y_{j+\frac{1}{2},m-\frac{1}{2}} \to 
\frac{F\,Y_{1,0}}{\sqrt{3}}\\
\psi_2&&=-\left[\frac{j+1+m}{2(j+1)}\right]^{1/2}F\,Y_{j+\frac{1}{2},m+\frac{1}{2}} 
\to \sqrt{\frac{2}{3}}\,F\,Y_{1,1}\nonumber\\
\psi_3&&=i\left[\frac{j+m}{2j}\right]^{1/2}G\,Y_{j-\frac{1}{2},m-\frac{1}{2}}\to
i\,G\,Y_{0,0}\nonumber\\
\psi_4&&=i\left[\frac{j-m}{2j}\right]^{1/2}G\,Y_{j-\frac{1}{2},m+\frac{1}{2}}\to
0.\nonumber
\ea
\noindent
Second solution:\\
\ba\psi_1&&=i\left[\frac{j+m}{2j}\right]^{1/2}L
\,Y_{j-\frac{1}{2},m-\frac{1}{2}}
\to i\,L\,Y_{0,0}\\
\psi_2&&=i\left[\frac{j-m}{2j}\right]^{1/2}L\,Y_{j-\frac{1}{2},m+\frac{1}{2}}\to
0\nonumber\\
\psi_3 &&= 
\left[\frac{j+1-m}{2(j+1)}\right]^{1/2}K\,Y_{j+\frac{1}{2},m-\frac{1}{2}} \to 
\frac{K\,Y_{1,0}}{\sqrt{3}}\nonumber\\
\psi_4&&=-\left[\frac{j+1+m}{2(j+1)}\right]^{1/2}K\,Y_{j+\frac{1}{2},m+\frac{1}{2}} 
\to \sqrt{\frac{2}{3}}\,K\,Y_{1,1}.\nonumber
\ea
The spherical harmonics are already normalized to one. The arrow indicates  that
the one and only value of the total angular momentum in conceptual connection with
the problem is
$j=1/2$. ($m=1/2$ is our option). The substitution of the sum  of these solutions in 
equation (\ref{eq1}) yields two identical systems of differential equations, each
containing only one pair of radial functions, $(F,G)$ or $(K,L)$. We may leave
functions $(K,L)$ aside for a while and work with functions $(F,G)$:
\bse\label{eq17}\ba \partial_r
F+\frac{2F}{r}-\frac{\lambda_e\,F}{r^2}\ =&&\left[\frac{E}{\hbar
c}-\frac{\alpha}{r}\right]\,G\\
\partial_r G-\frac{\lambda_e\,G}{r^2}\
=&&-\left[\frac{E}{\hbar c}-\frac{\alpha}{r}\right]\,F\ea\ese
The substitutions
\ba r&&=\frac{e^2}{E}\,s,\quad \alpha=\frac{e^2}{\hbar c},\quad
\beta=\frac{m_\nu}{m_e},\quad
\gamma=\frac{E}{mc^2},\nonumber\\
F&&=F\,^\dagger\,\exp(-\alpha^{-1}\beta\,\gamma\,s^{-1}),\nonumber\\
G&&=G\,^\dagger\,\exp(-\alpha^{-1}\beta\,\gamma\,s^{-1}),\nonumber\ea
reduce the system to            
\bse\label{eq18}\ba
s^{-2}\partial_s(s^2\,F\,^\dagger)=(1-s^{-1})\,\alpha\,G\,^\dagger\\
\partial_s\,G\,^\dagger=-(1-s^{-1})\,\alpha\,F\,^\dagger\ea\ese
The reader may verify that the two independent solutions,$(F,G)$ and $(f,g)$, can
be written as:
\ba
\sum_{k\,=\,0}^\infty{G_k^\dagger(\alpha,s)}=a_0+\sum_{k\,=\,1}^\infty{G_k
^\ddagger(s)\,\alpha^{2k}},\nonumber\\
\sum_{k\,=\,0}^\infty{F_k^\dagger(\alpha,s)}=\sum_{k\,=\,0}^\infty{F_k
^\ddagger(s)\,\alpha^{2k+1}}\\
\sum_{k\,=\,0}^\infty{f_k^\dagger(\alpha,s)}=b_0\,s^{-2}+\sum_{k\,=\,1}^\infty{f_k
^\ddagger(s)\,\alpha^{2k}},\nonumber\\
\sum_{k\,=\,0}^\infty{g_k^\dagger(\alpha,s)}=\sum_{k\,=\,0}^\infty{g_k
^\ddagger(s)\,\alpha^{2k+1}}
\ea

Concerning the first solution, function  $G_0^\dagger $ is the constant $a_0$, and
all the other functions as well as their first derivatives vanish at $s=1$.
Concerning the second solution, function $f_0^\dagger =b_0\,s^{-2}$, and all  the
other functions, as in the former case,  vanish in a smooth manner at the classical
electron radius. The procedure for the obtainment of solutions in powers of
$\alpha^2$ is iterative in nature and runs as follows. 

In the first solution, the first step is to insert  $G_0^\dagger $ into 
(18a). Integration is the second step. The constant of integration serves to force
the vanishing of  $f_0^\dagger$ at  $s=1$:
\ba s^{-2}\partial_s(s^2F_0^\dagger)=(1-s^{-1})\,\alpha\,a_0\nonumber
\\
\Rightarrow\qquad F_0^\dagger=\alpha(a_0/6)(s^{-2}-3+2s).\nonumber\ea    
We proceed with the insertion of $F_0^\dagger$ into (18b). Integration follows.
The constant of integration forces the vanishing of $G_1^\dagger$  at  $s=1$:
\ba
\partial_s\,G_1^\dagger=-(1-s^{-1})\,\alpha\,F_0^\dagger\qquad\Rightarrow\nonumber\\
G_1^\dagger=-\alpha^2(a_0/12)[s^{-2}-2s^{-1}+6\ln\,s+9-10s+2s^2]\nonumber\ea
and so on. Notice that function $G^\dagger$ cuts the $s$--axis only once at
$s^2\approx \alpha^2/12$. 

In the second solution, the iterative procedure starts
with the substitution of $f_0^\dagger=b_0\,s^{-2}$ into (18b), which yields
function $g_0^\dagger$: 
\ba \partial_s \,g_0^\dagger=-(1-s^{-1})\,\alpha\,b_0\,s^{-2}\nonumber\\
\Rightarrow\quad g_0^\dagger=-\alpha\,(b_0/2)\,(1-s)^2s^{-2}.\nonumber\ea     
The reader may  follow the prescription to get functions $f_1^\dagger$ and 
$g_1^\dagger$. With these functions at hand we may write down function
$g^\dagger\,G^\dagger\,s^{-1}$ which will be useful hereinafter:
\ba (g^\dagger\,G^\dagger\,s^{-1})\,s^2=
\alpha\,\xi_0+\alpha^3\,\xi_1+...+\alpha^{2k+1}\,\xi_k+...\ea
\ba\xi_0&&=a_0\,b_0\,[s^{-1}-2+s],\nonumber\\
\xi_1&&=-a_0\,b_0\,[s^{-3}-4s^{-2}+40-5s-36s^2\nonumber\\&&+4s^3+60s\ln\,s]/12\nonumber
\ea
Since functions: $F,\,g,\, gG$ and $fF$ vanish in a smooth manner at  $s=1$, the
physical properties of the electron should be represented with covariant densities
made up of the product of the two independent solutions: take for example the
4--vector
\ba 2\langle
\psi_a\gamma_\mu\psi_b\rangle=(\psi_a+\psi_b)\gamma_\mu(\psi_a+\psi_b)^*\nonumber
\\
-\psi_a\gamma_\mu\psi_a^*-\psi_b\gamma_\mu\psi_b^*\ea  
where $\psi_a$ and $\psi_b$ are the two independent solutions. 
The time component of this 4--vector,
\ba
\langle\psi_a\gamma_t\psi_b\rangle=\frac{1}{3}|Y_{1,0}|^2\,Ff+\frac{2}{3}
|Y_{1,1}|^2\,Ff+Y_{0,0}^2\,Gg\ea
vanishes in a smooth manner at 
$s=1$ ($\gamma_t$ is the unit matrix). However,  we cannot consider the
solution $\psi=0$ for the region  beyond  $s=1$; because doing so would make
functions $G$ and $f$ look  like strings with a loose end (see figure 1).

Graphical discontinuities in the wave functions are not acceptable. This problem is
solved by replacing the potential  $\alpha\,A_\mu$ in favor of      
$(m_e\,c\,\hbar^{-1})\,u_\mu $ for the region beyond the classical electron radius.
The corresponding radial equations are
\bse\ba
\partial_r\,F+\frac{2\,F}{r}-\frac{\lambda_e\,F}{r^2}&&= \ \frac{E-m_e\,c^2}{\hbar
\, c}\,G\\
\partial_r\,G-\frac{\lambda_e\,G}{r^2}&&= \ -\frac{E-m_e\,c^2}{\hbar
\, c}\,F\ea\ese 
Now we must set  $E=m_e\,c^2 $. With this consideration
the external solutions,
\ba G=a_0\,\exp(-\lambda_e\,r^{-1}),\qquad F=0,\\
f=b_0\,r^{-2}\,\exp(-\lambda_e\,r^{-1}),\qquad g=0,\ea
join with the internal solutions (19) and (20) at the classical
electron radius in a smooth manner. The external solutions have no further meaning 
since the corresponding bilinear products are null.

\begin{figure}
\centering
\includegraphics[height=15.2cm]{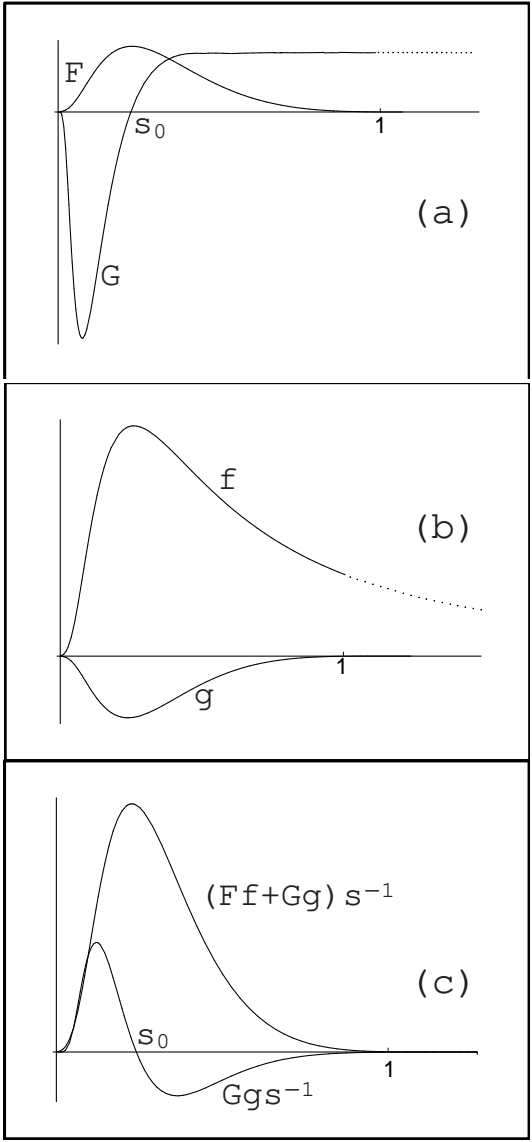}
%
%
\caption{The graphs of the
functions merely show their main features: all graphs vanish in a smooth manner at
the origin of coordinates. (a) In the first independent solution, function $F$ is
(roughly speaking) $\alpha$ times function $G$. Function $G$ cuts the $s$--axis only
once at  $s^2\approx\alpha^2/12 $. Function $F$ vanishes in a smooth manner at $s=1$. 
Beyond $s=1$, function $G$ is essentially constant.   (b) In the second independent
solution, function $g$ is about $\alpha$ times function $f$; the former smoothly
vanishes at $s=1$. Beyond this radius,  $f$ decays essentially as $s^{-2}$.   (c) The
products of the functions vanish in a smooth manner at $s=1$. Beyond the classical
electron radius the bilinear products are null. Thus, the energy of the electron is
confined within its classical radius.}
\label{fig:1}       
\end{figure}
           
The bilinear function (22) could be named interaction--density, the
recipient of the action of the Coulomb potential. Accordingly,
\begin{equation}{\cal E} \ = \
4\,\pi\,m_e\,c^2\,\int_0^1{A_\mu(s)\,\langle\psi_a\,\gamma_\mu\,\psi_b\rangle}
\,s^2\,ds\end{equation}
would represent the  electromagnetic inertia of the electron.
This definition, reminiscent of Maxwell's expression (see equation (4)) is the
formal substitution of the charge--density in favor of the interaction density. In
Maxwell's  theory the charge density determines the electrostatic potential,
whereas in the fundamental theory here being proposed the interaction density is
determined  by two  potentials endowed with independent existence.

\section{Further results}

The incorporation of the Dirac fifth matrix ($\gamma_5$ in equation (2)) into
this approach enables  one to write down the whole set of covariant densities
\cite{[3]}. 

The matrices,
\ba \gamma_t && = \ \textrm{diag}\,\left[1,1,1,1\right],\nonumber\\ \gamma_5 && = \
\textrm{diag}\,\left[1,1,-1,-1\right],\nonumber\\
\gamma_x\,\gamma_y && = \ \textrm{diag}\,\left[i,-i,i,-i\right],\nonumber\\
\gamma_x\,\gamma_y\,\gamma_5
&& = \ \textrm{diag}\,\left[i,-i,-i,i\right].\nonumber\ea
and the table below are particularly important in this regard. 
\begin{center}
\[
\begin{array}{ccc}
\textrm{Pseudo Invariant} & \qquad 1 & \qquad \textrm{none}\\
\textrm{Invariant}        & \qquad 1 & \qquad \left\langle {\psi _a\gamma _5\psi _b}\right\rangle\\
\textrm{4--vector}        & \qquad 4 & \qquad \left\langle {\psi _a\gamma _t\psi _b}\right\rangle\\
\textrm{pseudo 4--vector} & \qquad 4 & \qquad\left\langle {\psi _a\gamma _x\gamma_y\psi _b}\right\rangle\\
\textrm{anti--symmetric tensor}  & \qquad 6  & \qquad\left\langle {\psi _a\gamma _x\gamma_y\gamma _5
\psi _b}
\right\rangle
\end{array}
\]
\end{center}
\noindent
The first column shows the whole set of covariant densities. 
The second column indicates the number of components of each density. The third column
shows the only component of the density that survives volume integration; the other
components contain   products of different spherical harmonics and their volume
integral vanishes.\\

The transformation properties  of $S_z=\langle\psi _a\gamma _x\gamma_y \psi
_b\rangle $ and $M_z=\langle\psi _a\gamma _x\gamma_y\gamma _5
\psi _b\rangle $,
suggest to interpret the former as the z-component of spin-density  and 
the latter as the $z$--component of magnetic polarization. From the
matrices above and equations (15) and (16) we can see that $S_z$ is the
same for the two independent solutions, $(F;G,f,g)$ and  $(K,L,k,l)$.
Though, $M_z$, corresponding to the first solution is opposite to
$M_z$ of the second solution. Since electrons and positrons
with  the same spin orientation have opposite magnetic moments and the energy
of both solutions is positive,  perhaps the negative energy solutions
of the Dirac equation, as well as the philosophy built up around
them, need to be re--examined. Curiously enough, this is not the
only case where physicists have thought that  space itself has strange
properties: Special Relativity came to show the futility of the study of the
hydrodynamic properties the ether that presumably pervaded the whole universe.

   The graph of the density of the electromagnetic inertia  (see panel (c) of
figure 1) invites the volume integral of the function $gGs^{-1}$  to vanish.
This condition would interrelate $\alpha$ and $\beta$ and the electron  would
look as a minute planetary system wherein two electro-quarks (named so
because of the appearance of fractions $1/3$ and $2/3$ in (23)) retain all
the electromagnetic inertia while revolving around the singularity  of the
potentials. The covariant way of expressing this condition involves
the Dirac fifth matrix $\gamma_5$: the density  $\textrm{I}_1
=u_\mu\,\langle\psi _a\gamma _\mu\psi _b\rangle$ is an invariant and so
is $\textrm{I}_2 =\langle\psi _a\gamma _5\psi _b\rangle$. Therefore,
the vanishing of the volume  integral of \,
$\textrm{I}_0(\textrm{I}_1-\textrm{I}_2)$ is the condition we are
talking about. 

For the purpose of evaluating this integral it is important to notice
the following:\\

\begin{enumerate}
\item If $ s\gg 2\alpha^{-1}\,\beta=\eta$, function
$W=\exp(-\eta\,s^{-1})$ is close to $1$.

\item Function $W$ starts bending downward for values of $s$
comparable to $\eta$.

\item For $s\ll \eta$, function $W$ is essentially null.  

\end{enumerate}

If $\eta$ is sufficiently small we have:
\begin{equation}\int_0^1{W\,(1+s+s^{-2}+s^{-3})\,ds}\,\approx\,1+
\frac{1}{2}+\eta^{-1}+\eta^{-2}.\nonumber\end{equation} 
Now,
\begin{equation}\partial_\eta\,\int_0^1{W\,s^{-1}\,ds}
=-\eta^{-1}\,\exp(-\eta)\,\approx\,-\eta^{-1}+1-\eta/2+...
\nonumber\end{equation}
Therefore
\begin{equation}
\int_0^1{W\,s^{-1}\,ds}=-\ln\,\eta+\eta-\eta^2/2!\,+...+c_0\end{equation}
The constant of integration $c_0\approx -0.51$ can be evaluated  by 
plotting  the graph of \, $W\,s^{-1}$ (for $\eta=1$) and comparing the
area under the graph with its analytical value given by the right hand
side of (28) (evaluated for $\eta=1$).  

When $\eta$ is small enough, the dominant term of the integral of
$W\,\xi_1$ (see equation (21)) is the first one.
All these considerations imply that the vanishing of the volume 
integral of $G\,g\,s^{-1}$ is congruent with the following approximation
\begin{equation}-\ln\,\eta -
0.51-2+1/2 \ = \ (\alpha^2/12)\,\eta^{-2}.\end{equation}
Considering that \, $ m_e=511000$ eV and that $\alpha=1/137$, we get
that the mass of the e--neutrino is larger than $1.75$ eV but 
smaller than $1.78$ eV. Hitherto experimental  physics  has  set  an
upper limit of $2.2$ eV for $m_\nu$ \cite{[6]}.

\section{The electron--neutrino solution}

Equation (1) provides an amazingly simple solution to the electron--neutrino 
if it is regarded as an electron deprived of
Coulomb potential. In this case the self--action would take this form:
\begin{equation}S_\mu(\nu) \ = \ -m_\nu\,c\,
\hbar^{-1}\,u_\mu+i\,\lambda_e\,\partial_\mu\,\textrm{I}_0\end{equation} 
We already know the corresponding radial equations, yet there is only one
solution of interest:
\begin{equation} F \ = \ s^{-2}\,\exp(-\beta^2\,s^{-1}),\qquad G
\ = \ 0\end{equation}
The 
fact that it is not possible to cut off the solution at any particular
radius suggests the Copenhagen interpretation of the square of the wave
function:  the probability of the electron--neutrino to interact with matter
beyond one wave length from the singularity of its own potential is
\ba
\left(\int_1^\infty{u_\mu\,\psi\,\gamma_\mu\,\psi^*\,s^2\,ds}
\right)\,\left(\int_0^\infty{u_\mu\,\psi\,\gamma_\mu\,\psi^*\,s^2
\,ds}\right)^{-1}\nonumber\\
= \ 1-\exp(-2\,\beta^2) \ \approx \ 2\,(m_\nu/m_e)^2 \ < \ 10^{-10}
\nonumber
\ea                                                                                                                                          
Neutrinos $\mu$ and $\tau$ could be described with other coupling
constants, ($\lambda_\mu$ and $\lambda_\tau$).\\

\section{The proton}

The description of the proton can be carried out thinking of it as a positron
endowed with a Yukawa neutral meson field. Consider the scalar equation
\begin{equation}  \partial_\alpha\,\partial_\alpha\, \zeta \ = \
(m_{\pi_0}\,c\,\hbar^{-1})^2\,\zeta,
\end{equation}
its time independent solution
\begin{equation}\zeta \ = \
r^{-1}\exp[-(m_{\pi_0}\,c\,\hbar^{-1})\,r]\nonumber
\end{equation} 
was used by Yukawa in his
meson theory of the nuclear force~\cite{[5]}. In general, the nuclear force is
attractive beyond the proton radius and repulsive for shorter  distances, which
prevents the nucleus from collapsing. Accordingly, the simplest self--action
that we can think of is 
\begin{equation}S_\mu(p) \ = \
-n\,\zeta\,u_\mu-\alpha\,A_\mu+i\,\lambda_e\,\sigma_\mu,
\end{equation}
where $n$ is a dimensionless coupling constant which must vanish when the
meson mass vanishes. The spin potential is assumed to be the same for the
electron, the neutrino and the proton, all the offspring of neutron decay. 

The solution of equation (1) for this case can be obtained following
procedures very similar to the one used for the electron case. Moreover, it
is not necessary to use numerical analysis for the obtainment of a good
approximation when $n$ is one of the two simplest
fundamental expressions: $n=m_{\pi_0}/m_p\approx 1/7$,\, or
$n=m_{\pi_0}\,m_p/(m_{\pi_0}+m_p)^2\approx 1/9$. Actually, the correct value
of \, $m_e/m_p$ \, is obtained with the vanishing of the volume integral of
$G\,g\,(\alpha\,\textrm{I}_0+n\,\zeta)$ when $n\approx 1/11$. Therefore, it 
can be inferred that the proton self--energy was defined as
\begin{equation} {\cal
E} \ = \ 4\pi\,m_pc^2\int_0^{s_0}{(\alpha\,A_\mu+n\,\zeta\,u_\mu)\,
\langle\psi_a\,\gamma_\mu\,\psi_b\rangle\,s^2\,ds}\end{equation}
where $s_0$ is the radius where the right hand side of the corresponding
radial equation vanishes.

This model can be improved to get a better approximation of $m_e/m_p$.
However, we see a drawback of a different kind: the prevailing consensus is
that protons are made up of three spin $1/2$ subparticles. Perhaps an
attempt to express the GUT's model in a covariant way should be made. How?
assigning to each quark an individual bispinor wave function, yet all of
them interrelated in a system of differential equations similar to (1).    

One more interesting aspect of scalar equations is that the substitutions
\,$
\psi(\vec{\textrm{r}},t)=\psi(\vec{\textrm{r}})\,\exp[-i\,(m\,c^2+E)\,
t\,\hbar^{-1}]$ and $A_\beta\,\partial_\beta=c^{-1}\,\phi\,\partial_t $ in
\begin{equation}
(\partial_\beta\,\partial_\beta-2\,i\,\alpha\,A_\beta\,
\partial_\beta)\,\psi \ = \ (m^2\,c^2\,\hbar^{-2})\,\psi\end{equation}  
yield
\ba &&-(\hbar^2/2m)\, \nabla^2\,\psi +
e^2\,\phi\,[1+(E/mc^2)]\,\psi \nonumber\\ &&= \ E\,\psi\,[1+(E/2mc^2)].
\ea 
Neglecting $E/mc^2 $, which in atomic physics is much
smaller than self--energy, the familiar Schr\"{o}dinger equation is obtained.
However, the important point is that imaginary vector potentials are necessary to
yield differential equations with real coefficients. This is a feature of
relativistic wave equations with the time--dependent factor
$\exp(-i\,t\,\hbar^{-1})$. In fact, we had this situation with equation (1) and
the potential\,
$i\,\lambda_e\,\sigma_\mu$.

As a final remark, the time--dependent factor $\exp\,(-E\,t\,\hbar^{-1})$ in
equation (1) is a logical step to start investigating spin $1/2$ unstable
particles.\\  

\section*{Acknowledgments}
This work was finished at the Instituto 
de  Ciencias Nucleares,  Universidad Nacional
Aut\'onoma de M\'exico (UNAM). I am grateful to doctors Octavio  Casta\~nos,
director of the institute, and Alejandro Frank, head of the Department of
Nuclear Physics, for their hospitality and kind attention. I appreciate
very much the interesting and helpful comments of Dr. Roberto Sussman.

\end{document}